\documentclass[final,5p,times,twocolumn]{elsarticle}
\usepackage{amssymb}
\usepackage{graphicx}
\usepackage[caption = false]{subfig}
\journal{Physics Letters B}

\begin{document}

\begin{frontmatter}

\title{Phase transition in compact stars due to a violent shock}

\date{\today}

\author[fias,nrcki]{Igor Mishustin}
\ead{mishustin@fias.uni-frankfurt.de}

\author[fias]{Ritam Mallick}
\ead{mallick@fias.uni-frankfurt.de}

\author[fias]{Rana Nandi}
\ead{nandi@fias.uni-frankfurt.de}

\author[fias,nrcki]{Leonid Satarov}
\address[fias]{Frankfurt Institute for Advanced Studies, 60438 Frankfurt am Main, Germany} 
\address[nrcki]{National Research Center “Kurchatov Institute”, Moscow 123182, Russia}
 
\begin{abstract}
In this letter we study the dynamics of a first order phase transition from nucleonic to quark matter in neutron stars.
Using standard equations of state for these two phases we find the density range where such a transition is
possible. Then we study the transformation of the star assuming that the quark core is formed via a spherical 
shock wave. The thermodynamical conditions in the quark core are found from the conservation laws across the transition
region. Their dependence on the density and velocity of the incoming nuclear matter are studied. It is found that the shock is
especially violent in the beginning of the conversion process when the velocity of the infalling matter is especially high. 
As the shock propagates further from the center the front velocity first increases and reaches a maximum value when the incoming 
velocity is around $0.2$. Finally, the front velocity 
quickly goes to zero when incoming matter velocity approaches zero. We have shown that the density and pressure jumps are especially large in the 
begining of the transition process.
\end{abstract}


\begin{keyword}
dense matter, stars: neutron, equation of state
\end{keyword}

\end{frontmatter}


The most interesting property of dense baryonic  matter is its possible phase transition (PT) to the deconfined phase
at supranuclear densities, a few times the nuclear equilibrium density $n_0=0.15$ fm$^{-3}$. Transformations
of neutron star (NS) to quark star (QS) or to pion- or kaon-condensed star shave been 
studied in Refs. \cite{igor,haensel82,banik01,gentile93,igor2,mallick06,mallick07,drago08}. However, since the NS interiors 
are beyond the direct observation, the conclusions strongly depend on models used to describe such 
PT. The PT from NS to QS is presumably a much more violent process than the transition to a pion or
kaon condensed star. In case of a first order PT there may be even a possibility that the whole star is 
converted to a QS, which is normally a strange quark star (SQS) due to a large fraction of strange 
quarks. The phase transformation is usually assumed to begin at the centre of a star when the density increases beyond the critical 
density. The PT can be triggered by several processes: slowing down of the rotating star \cite{glen}, accretion
of matter on the stellar surface \cite{alcock} or simply cooling. It is most likely that the PT happens due to
the nucleation process near the star centre.
Such a PT should be accompanied by significant energy release in the form of latent heat, which leads to a neutrino burst thereby cooling the star. 
This energy release should have several observable signatures like the gamma ray bursts
\cite{bombaci,mallick} and change in the cooling rate. It will be interesting to study possible manifestation
of such strong PT under realistic astrophysical conditions.

We assume that the dynamics of the PT is characterized by several stages: the transition process may begin suddenly due to a fluctuation in
the star density. The PT should start after the central density exceeds some critical value. Then the region of the new phase 
will propagate to the periphery. The transition can be of 
quasi-equilibrium type as in slow burning, or may be more violent occurring via a shock front. Since there is no way
to stop the propagation of the shock at the equilibrium position in the star, it will overshoot the point and gradually slow down until it
finally stops. Then the reverse process may start when the front moves in the opposite direction converting 
the quark matter (QM) again to hadronic matter. Apparantely, this process will later stop after some again, and the PT front will 
oscillate around its equilibrium position.

\begin{figure}
\includegraphics[width = 250pt]{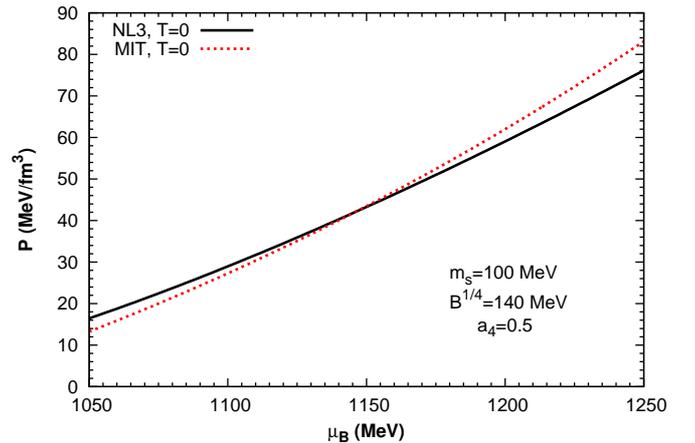}
\caption{(Color online) Pressures of nucleonic and quark matter as functions of baryon chemical potential at zero temperature. The model parameters for
the quark matter are shown in the figure. The intersection point corresponds to the equilibrium PT from nucleonic to
quark matter.}
\label{PvsmuB_FT}
\end{figure}

As already mentioned, the PT brings about significant energy change in the star. At least a part of this energy will be transformed 
into internal heat. Therefore, even if the nuclear matter was initially cold, the produced  QM should have
finite temperature. The actual temperature of the shocked matter will depend on the equations of state (EoS) of
two phases. For the hadronic phase we adopt the relativistic mean field approach which is widely used to describe
the hadronic matter in compact stars. The corresponding Lagrangian can be represented in the following form
\cite{serot86,boguta,glendenning} ($\hbar =c=1$)

\begin{eqnarray} 
& {\cal L}_H = \sum_{n} \bar{\psi}_{n}\big[\gamma_{\mu}(i\partial^{\,\mu}  - g_{\omega n}\omega^{\,\mu} - 
\frac{1}{2} g_{\rho n}\vec \tau . \vec \rho^{\,\mu})- \nonumber \\
& \left( m_{n} - g_{\sigma n}\sigma \right)\big]\psi_{n} + \frac{1}{2}({\partial_{\,\mu} \sigma \partial^{\,\mu} \sigma - m_{\sigma}^2 \sigma^2 } )
\nonumber\\
& -\frac{1}{3}b\sigma^{3}- \frac{1}{4}c\sigma^{4} - \frac{1}{4} \omega_{\mu \nu}\omega^{\,\mu \nu}+ 
\frac{1}{2} m_{\omega}^2 \omega_\mu \omega^{\,\mu} \nonumber \\ 
& -\frac{1}{4} \vec \rho_{\mu \nu}.\vec \rho^{\,\mu \nu} + \frac{1}{2} m_\rho^2 \vec \rho_{\mu}. \vec \rho^{\,\mu} 
+ \sum_{l} \bar{\psi}_{l}    [ i \gamma_{\mu}  \partial^{\,\mu}  - m_{l} ]\psi_{l}. 
\label{baryon-lag} 
\end{eqnarray}

\begin{figure*}
\captionsetup[subfigure]{labelformat=empty}
\subfloat[]{\includegraphics[width = 250pt]{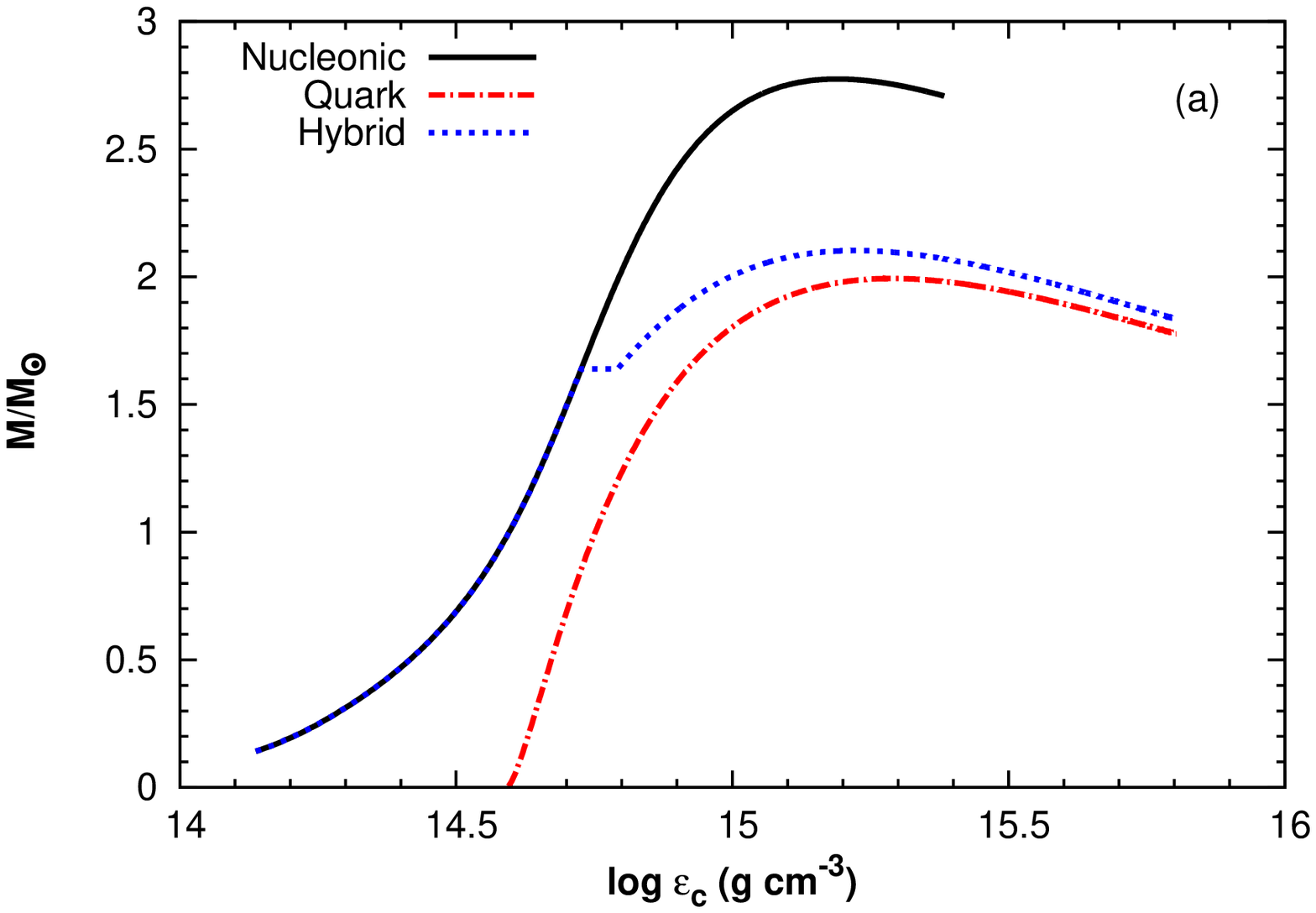}} \quad
\subfloat[]{\includegraphics[width = 250pt]{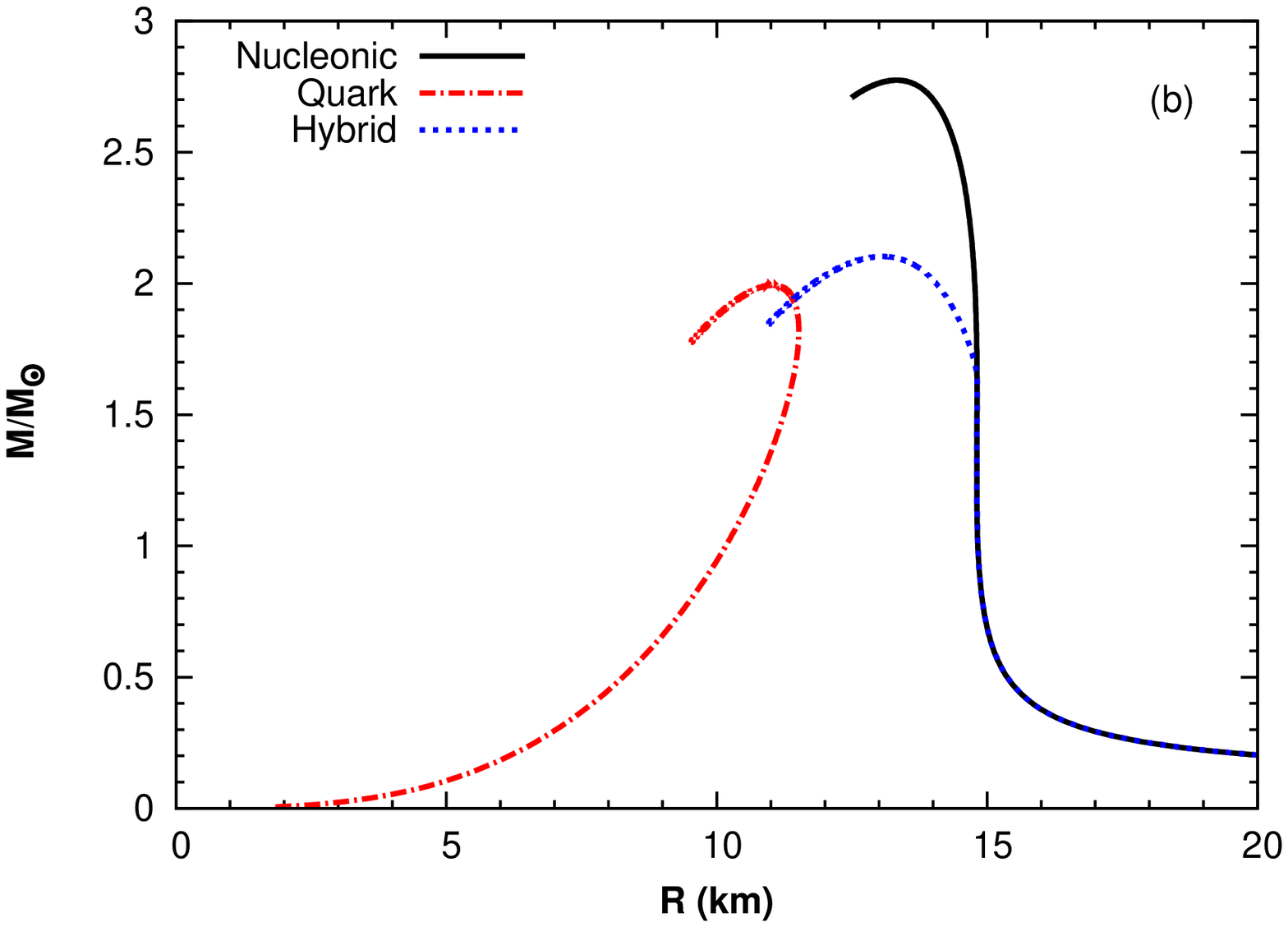}}
\caption{ Mass as functions of central energy density (left panel) and radius (right panel) for pure nucleonic (solid lines), pure quark (dashed curves)
and hybrid stars (dotted lines). All calculations are made for $T=0$.}
\label{mr}
\end{figure*}

In our calculation we take into account only nucleons ($n$) and leptons ($l=e^{\pm},\mu^{\pm}$)
The latter are assumed to be non-interacting, but the nucleons are
coupled to the scalar $\sigma$ mesons, the isoscalar-vector $\omega_\mu$ mesons and the isovector-vector 
$\rho_\mu$ mesons. The adjustable constants of the models are fitted to reproduce basic properties of nuclear 
matter and finite nuclei. We use the NL3 parameter set \cite{nl3} which generates sufficiently massive NS (see Fig. 2). 

To describe the QM we apply the MIT bag model \cite{chodos} which was already used earlier to describe the
SQS \cite{alford}. This model is able to generate stars up to $2.3$ solar mass if one includes quark
interactions \cite{weissenborn11}. The model is defined by the grand potential
\begin{equation}
 \Omega_Q=\sum_i \Omega_i +\frac{\mu^4}{108\pi^2}(1-a_4)+B
\end{equation}
where $i$ stands for quarks and leptons, $\Omega_i$ denotes the grand potential for an ideal gas of species $i$,
$B$ is the bag constant and the second term accounts for the interaction of quarks. Here $\mu$ is the baryon chemical potential, 
the quark interaction parameter $a_4$ is
varied between 1 (no interaction) and 0 (full interaction). 
We take into account the $u,d$ and $s$ quarks. It is assumed that the masses of $u$ and $d$ quarks are equal to $5$ and $10$
MeV respectively, the mass of $s$ quark is taken to be $100$ MeV. In our calculation we choose the values of $B^{1/4}=140$ MeV and $a_4=0.5$. 
With such parameters we are able to satisfy the constraint imposed by the observation of $2$ solar mass
pulsars (see below). In this model it is also quite easy to
include finite temperature effects. 

Let us first analyse the equilibrium PT between the nucleonic and quark phases at $T=0$. Fig. \ref{PvsmuB_FT} shows  
pressures of the two phases as functions of baryon chemical potential. The intersection of the curves at $\mu_c \simeq 1150$ MeV 
determines the point of PT from cold nucleonic to cold quark matter. We have solved the TOV equations to find the equilibrium star 
sequences for the considered EoSs. The PT is implemented assuming the Maxwell construction \cite{bhat}.
The results are presented in Fig. 2. As one can see, the PT changes the hydrostatic equilibrium 
in the star and gives rise to a new branch of hybrid stars. 
These stars have lower masses than the NS at the same central energy density (Fig. 2a).
For nucleonic stars the maximum mass is about 2.8$M_{\odot}$, whereas for hybrid star we obtain the value 2.1$M_{\odot}$, still within 
the present constraints of 2$M_{\odot}$ \cite{demorest,antonidis}. The radius at the maximum-mass for hybrid stars is about $13$ km,
slightly smaller than that of NS (Fig. 2b).

\begin{figure}
\captionsetup[subfigure]{labelformat=empty}
\subfloat[]{\includegraphics[width = 250pt]{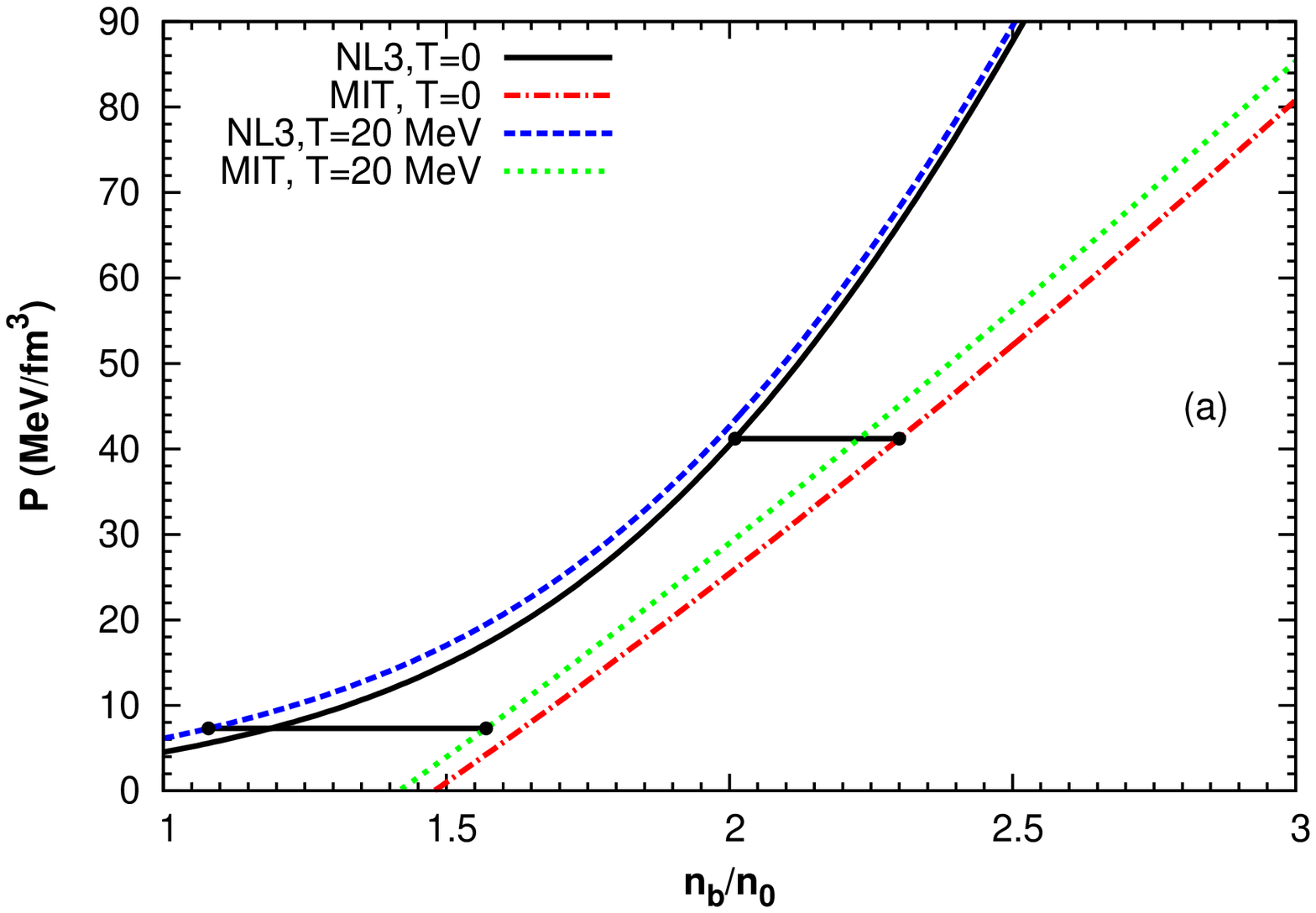}} \quad
\subfloat[]{\includegraphics[width = 250pt]{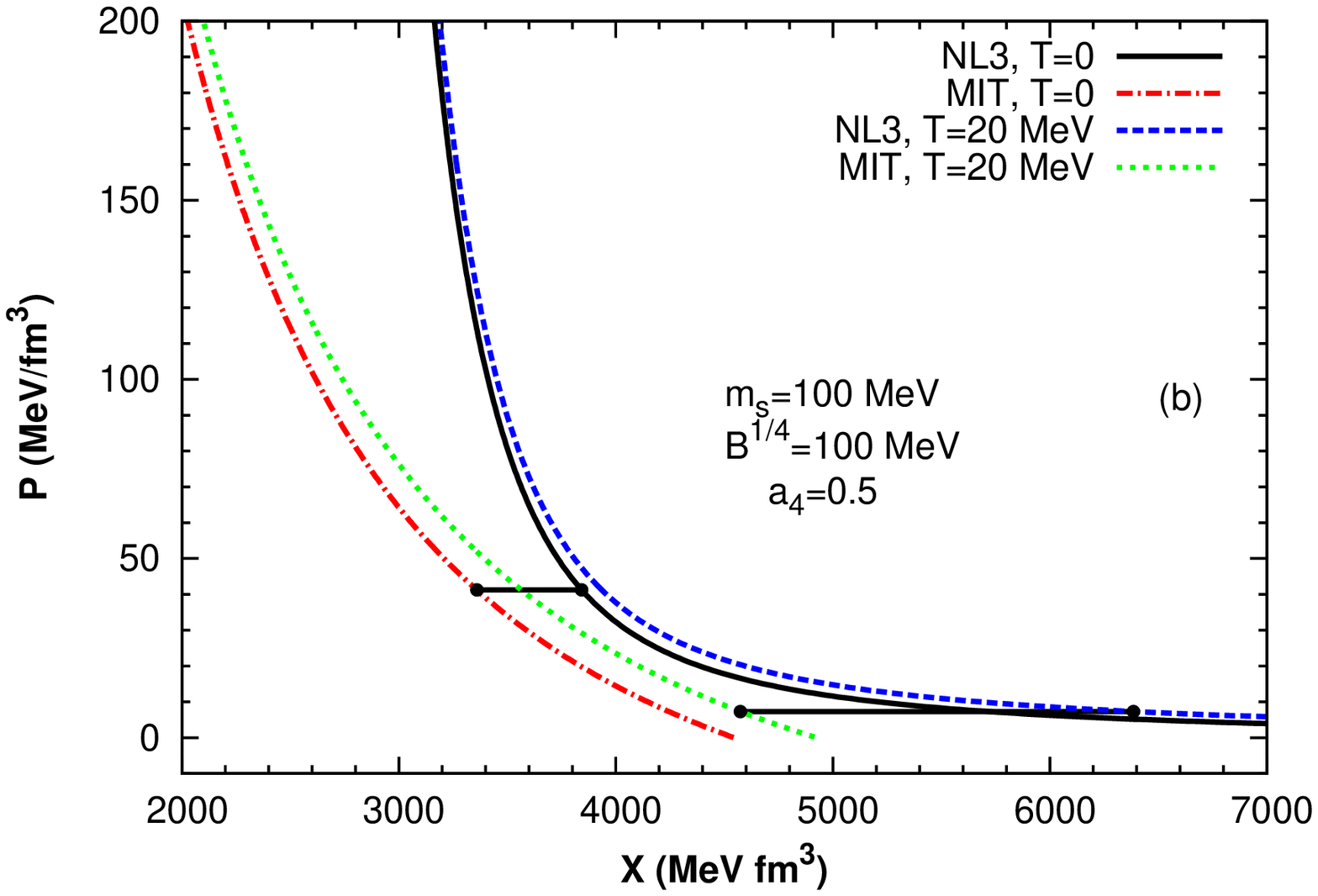}}
\caption{a) Pressure isotherms as functions of normalized baryon density.
b) The Poissons adiabats plotted in the $X-P$ plane both for nucleonic and quark matter. 
Solid black lines indicate baryon and energy density jumps corresponding to
equilibrium PTs.}
\label{PvsX}
\end{figure}

At nonzero temperatures it is convenient to represent the EOS in the form of the pressure isotherms. In  
Fig. \ref{PvsX}a we show some of them as functions of baryon density. The PT is represented by a horizontal line connecting two phases at 
densities $n_i=(\partial P/\partial \mu)_{\mu_c}$ ($i=q,n$), where $\mu_c$ is the chemical potential corresponding to the crossing point in Fig. 1.
In Fig. 3b we show the pressure isotherms as functions of 
$X \equiv (\epsilon+p)/{n_b}^2$ which is more appropriate for our future analysis. In the nonrelativistic limit 
$X \simeq M_N/n_b$, where $M_N$ is the nucleon 
mass. As the temperature increases the curves shift to higher $X$ which correspond to less dense matter.
The $X$ values decrease but $n_b$ grows as we go to the center of a star.
For the nucleonic matter (NM) the value of $X$ is larger than for the QM, because for the same values of pressure the 
baryon density of QM 
is higher. Therefore, during the PT from the nucleonic
to quark phase, the values of $X$ and baryon density exhibit jumps, which become stronger at larger initial densities. 
For zero temperature in a static matter the PT is predicted at density $n_b \simeq 2n_0$ while at $T=20$MeV the PT occurs at lower density 
($n_b \eqsim 1.1n_0$).
In a dynamical environment the PT from NM to QM 
may be delayed due to the barrier separating the two phases and therefore it occurs at higher density (see below).  
Moreover, since the PT from NM to QM leads to jumps in thermodynamical quantities, one may 
expect the formation of a step-like spatial profiles like in a shock wave. We assume that a shock-like discontinuity is generated 
somewhere near the centre of a spherically-symmetric star and later on it propagates outwards, leaving
behind a compressed quark core.  It is also assumed that the width of the transition zone is small in comparison to its 
radius and the matter flow is purely radial. 

The physical quantities in the nucleonic 
and quark phases (denoted below by subscripts $n$ and $q$, respectively) on both sides of the front
are related by the energy-momentum conservation and baryon number conservation laws. In the rest frame
of the front one can write the equations \cite{taub,landau,zeldovich}
\begin{eqnarray}
 w_n\gamma_n^2v_n=w_q\gamma_q^2v_q ,\label{econsv}\\
 w_n\gamma_n^2v_n^2+p_n=w_q\gamma_q^2v_q^2+p_q ,\label{pconsv}\\
 n_nv_n\gamma_n=n_qv_q\gamma_q \label{nconsv}.
\end{eqnarray}
Here $v_i$ ($i=n,q$) are the flow velocities, $p_i$ are the pressures, $w_i=\epsilon_i +p_i$ are the specific enthalpies and 
$\gamma_i=(1-{v_i}^2)^{-1/2}$ are the Lorentz-factors. 
Solving Eqs.~(3)--(4) with respect to $v_n,v_q$ one gets the expressions 
\begin{eqnarray}
|v_n|=\left[\frac{(p_q-p_{ n}) (\varepsilon_q+p_n)}{(\varepsilon_q-\varepsilon_n) (\varepsilon_n+p_q)}\right]^{1/2}, \nonumber \\
~~|v_q|=\left[\frac{(p_q-p_{ n}) (\varepsilon_n+p_q)}{(\varepsilon_q-\varepsilon_n) (\varepsilon_q+p_{ n})}\right]^{1/2},
\label{evel}
\end{eqnarray}
where $\varepsilon_i$ is the energy density of the $i$--th phase ($i=n,q$). Substituting (\ref{evel}) into~Eq.~(5)
gives the so-called Taub adiabat (TA) which is the equation connecting $\varepsilon_i,p_{ i},n_i$ on both sides of the shock
front:
\begin{equation}
 (p_{n}+\varepsilon_q) X_q=(p_q+\varepsilon_n) X_n\,,
 \label{eqta}
\end{equation}
where $X_i=w_i/n_i^2$. If one knows the EoS of the $i$-th phase, $p_{i}=p_{ i}(\varepsilon_i,n_i)$, the 
thermodynamic quantities $\varepsilon_i,n_i$ can be regarded as functions of $X_i,p_{i}$\,. For a given initial state of 
the NM, one can represent the TA of the QM by a line in the \mbox{$X_q-p_q$} plane. In general, this line does 
not go through the point with coordinates ($X_n,p_{n}$). The slope of the ''Rayleigh'' line, connecting this initial point with 
the point $(X_q,p_q)$ on the TA is proportional to $(\gamma_n v_n)^2$ \cite{landau}. On the other hand, for each velocity of incoming matter $v_n$
there is a~specific point on the TA corresponding to state of compressed quark phase.  

One can get simple explicit relations for strong shocks, i.e. in the limit $p_q\to\infty$\,. Neglecting in~Eqs. (\ref{evel})--(\ref{eqta})
 $\varepsilon_n,p_{n}$ as compared to  $\varepsilon_q,p_q$\,, one obtains the approximate relations: 
 \begin{equation}
   |v_n|\simeq 1,~~|v_q|\simeq X_q/X_n\simeq p_q/\varepsilon_q\,.
   \label{veta}
 \end{equation}
Assuming that at high energy densities the QM has the ultrarelativistic EoS, $p_q\simeq\varepsilon_q/3$, we get at
$p_q\to\infty$
\begin{equation}
 |v_q|\simeq 1/3\,,~~X_q\simeq X_n/3\,.
 \label{veta1}
\end{equation}

To estimate the velocity of the incoming nuclear matter $v_n$ one can use the continuity equation for the baryon current in the 
incoming NM outside the quark core. In the case of spherical symmetry it reads as
\begin{equation}
 \frac{\partial (n\gamma)}{\partial t} + \frac{1}{r^2}\frac{\partial}{\partial r}(n\gamma vr^2)=0 ,
\end{equation}
where, $v(r)$ is the radial velocity.
In the approximation of an incompressible fluid, ignoring the first term we get $\gamma vr^2\approx$ const.

The dynamics of the star transformation depends strongly on the density ratio at the phase boundary 
$\lambda=n_q/n_n$. As shown in ref. \cite{ramsey} the star is unstable for $\lambda>3/2$. In this case a
macroscopic fraction of the star transforms rapidly into a new phase. However, the dynamics of 
the star transformation also depends essentially on the velocity of the shock front created by the PT. 
To demonstrate this we solve Eqs. (\ref{econsv})-(\ref{nconsv}) taking
the baryon density, pressure and the incoming flow velocity of the NM as inputs. Using these equations, we find the 
density, pressure and velocity $v_q$ of the quark phase. At known quark density one can calculate $\lambda$. 
Now, we go from the front rest frame to the global frame, which is the reference frame where the QM is at rest. In this global frame 
the NM moves toward the center with velocity $\tilde{v}_n=(v_n+v_f)/(1+v_nv_f)$, where $v_f$ is the front velocity given by $v_f=-v_q$.
The values of $\lambda$ and $v_f$ fully determine the dynamics of the star transformation.

\begin{figure}
\includegraphics[width=250pt]{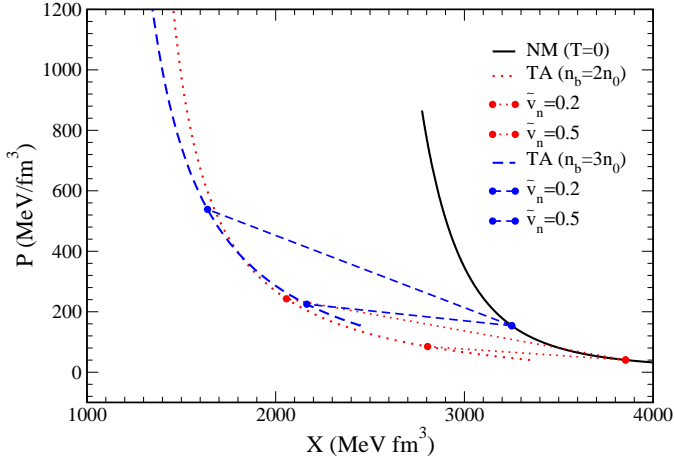}
\caption{The Taub adiabats for shock transitions from nucleonic to quark phase at different initial baryon densities $n_n=2n_0$ (dotted line) 
and $n_n=3n_0$ (dashed line). Lines with markers corresponds to shocks with different values of $\tilde{v}_n$. The solid line shows the zero temperature 
isotherm of nucleonic matter.}
\label{taub}
\end{figure}

In general the PT does not follow the equilibrium scenario and occurs with a certain degree of overshooting. Some
additional compression of NM is required to trigger the nucleation mechanism or reach the point of 
spinodal instability (see recent discussions in Ref. \cite{buballa14}). This will make the shock transition even more
violent and happening at higher densities, as compared to Fig. \ref{PvsX}a. Below we study such a ``delayed'' PT at 
densities $n_b=2n_0$ and $3n_0$, choosing various flow velocities of the incoming NM. 
We assume that close to the center of the star the velocity of incoming matter should be very high
($\gamma_nv_n \propto 1/r^2$).

\begin{figure}
\includegraphics[width=250pt]{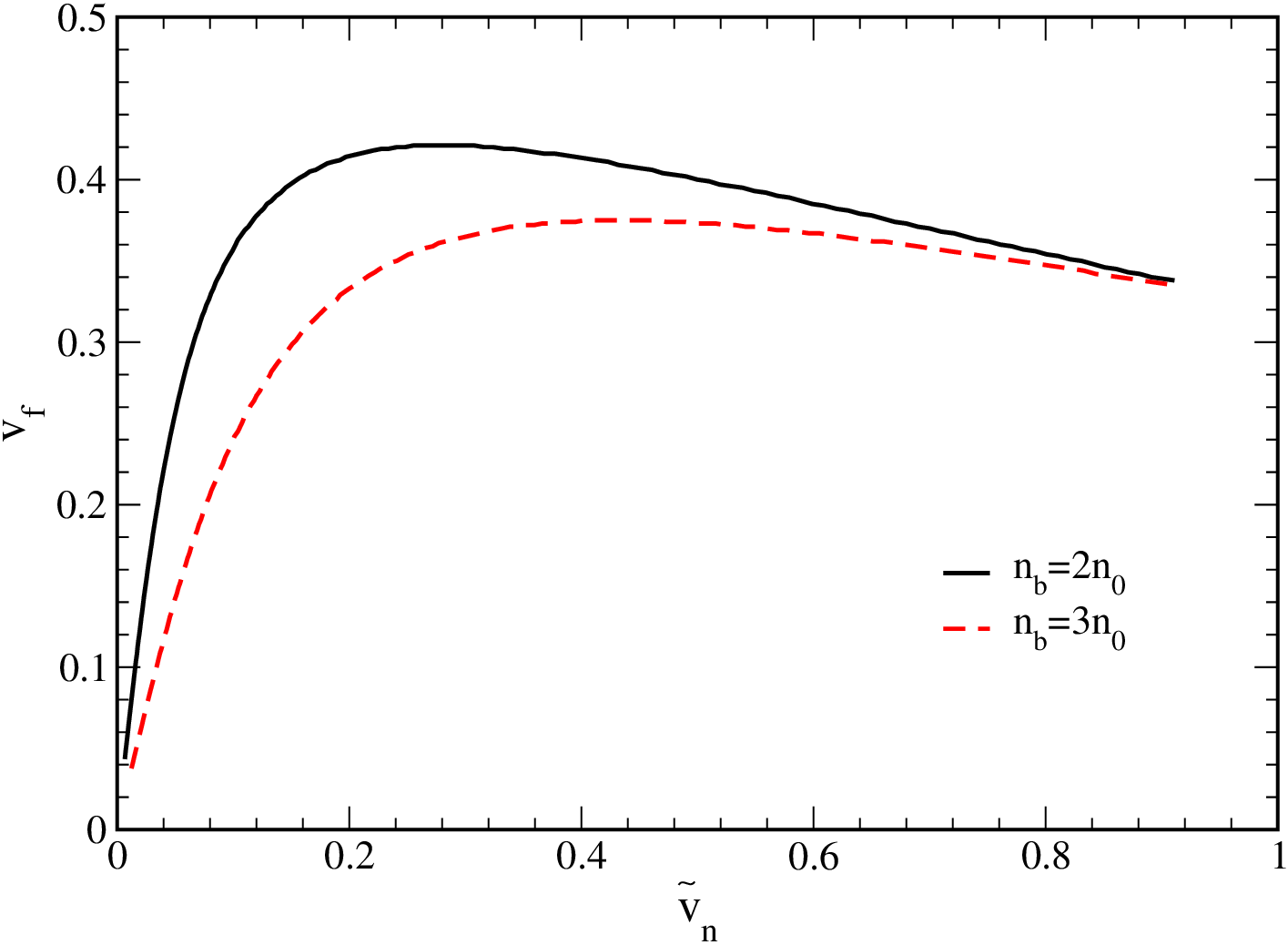}
\caption{Front velocity $v_f$ as a function of incoming NM velocity $\tilde{v}_n$ for two initial baryon densities.}
\end{figure}

Figure \ref{taub} shows the set of Taub adiabats of QM starting from different initial states of NM. 
One can see that indeed the final QM state strongly depends on the flow velocity and the density of incoming nucleonic
matter from Fig. 4. One can make the following conclusions: First, when the 
incoming flow velocity increases, the pressure and density of the final state increases too; 
Second, at a given onset density the pressure jump across the front increases with
the flow velocity of incoming matter.

The dynamics of the discontinuity can be better understood when we plot the front
velocity ($v_f$) as a function of incoming matter velocity ($\tilde{v}_n$) as shown in Fig. 5.
As was explained above, near the centre of the star the velocity of the incoming matter is largest according to the relation 
$\gamma_nv_n \propto 1/r^2$. Therefore, to understand the variation of the front velocity with time one should move from
right to left along the curves. Initially $\tilde{v}_n \simeq 1$ and $v_f \simeq 1/3$ which is in agreement with Eqn. (\ref{veta1}) 
valid for strong shocks. 
As the shock wave propagates outwards from the center the front velocity increases gradually. However, after $\tilde{v}_n$ becomes 
less than about$0.2$, the front velocity drops rather quickly and vanishes at $\tilde{v}_n \rightarrow 0$. 
A more detailed information concerning the shock properties for $\tilde{v}_n=0.2$ is shown in Fig. \ref{taub}.
It is interesting to note that the 
conservation conditions 
act in such a way that without any dissipation mechanism there is some type of deceleration which drives the front velocity to zero 
at some point inside the star, corresponding to static configuration with PT. 

Although the front velocity vanishes, the acceleration 
may not be zero and the front may overshoot this equilibrium position where $\Delta P \equiv P_q-P_n=0$. At later times the front will propagate inwards 
and quark matter will transform to nuclear matter in the rarefaction wave.
It may oscillate for some time around the zero value and ultimately the shock will stop due to dissipation.  

\begin{figure}
\includegraphics[width=250pt]{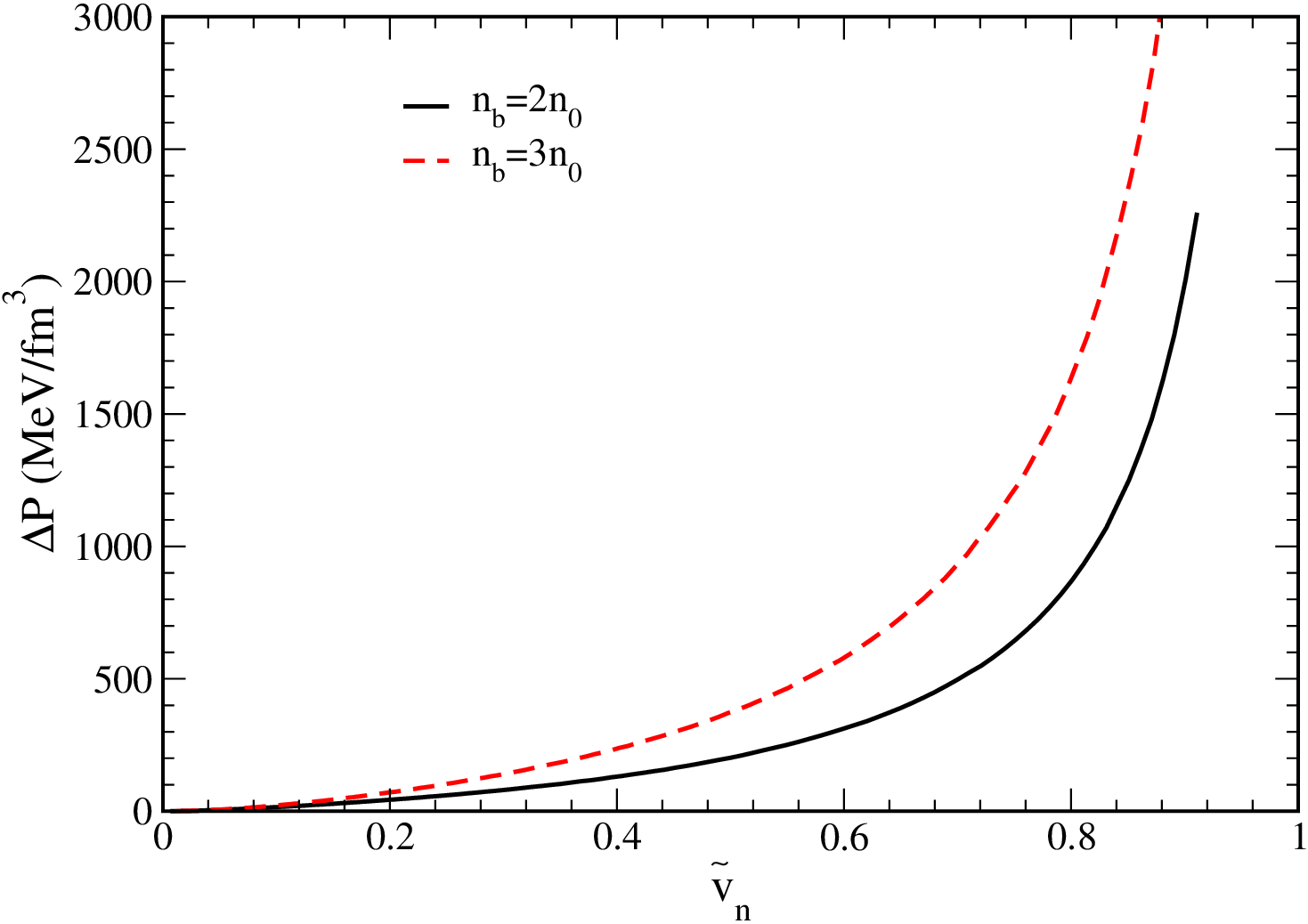}
\caption{Pressure jumps ($\Delta P = P_q - P_n$) across the PT front as a function of $\tilde{v}_n$ for two initial onset densities.}
\label{Pjump}
\end{figure}
 
In Fig. \ref{Pjump} we plot the pressure jump across the shock front as a function of $\tilde{v}_n$.
Near the center where the incoming matter velocity is large, the pressure jump
is also high, i. e. we have a strong shock induced PT.
In this region the curves are very steep signalling large pressure jump across the front. As $\tilde{v}_n$ goes down 
the pressure difference across the front diminishes and finally vanishes when $\tilde{v}_n \rightarrow 0$.

\begin{figure}
\includegraphics[width=250pt]{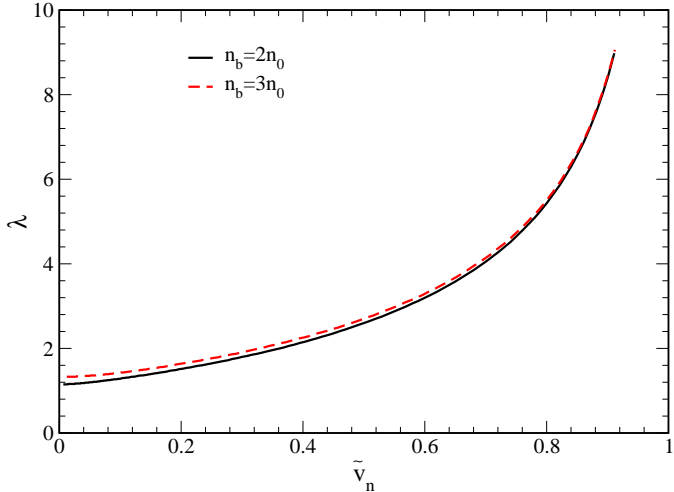}
\caption{Same as in Fig. 6 but for the density ratio $\lambda$.}
\label{lambda}
\end{figure}

In Fig. \ref{lambda} we plot the density ratio $\lambda=n_q/n_n$ as a function of $\tilde{v}_n$. Near the center, where
the incoming matter velocity $\tilde{v}_n$ is largest, the density jump across the front is largest too.
One can see that in this case we get values of $\lambda$, well above the critical value $\lambda=3/2$
for which the static star becomes unstable \cite{ramsey}. 
As $\tilde{v}_n$ decreases, $\lambda$ decreases gradually and reaches the values predicted by equilibrium PT at $\tilde{v}_n \rightarrow 0$.
For our choice of EoS, $\lambda$ becomes
less than $3/2$ at $\tilde{v}_n \lesssim 0.2$. 
Apparently, this means that the star approaches an equilibrium state with a relatively small core. 

According to Fig. 6, at fixed $\tilde{v}_n$ the pressure jumps across the front are 
different for different onset densities (one can see that it is larger for onset density $n_b=3n_0$). However, the values of 
$\lambda$ only weakly depends on the 
onset density and the curves for $n_b=2n_0$ and $n_b=3n_0$ almost coincide when $\tilde{v}_n \gtrsim 0.7$. This points to the conclusion that 
the shock transition is likely to be very violent as long as $\tilde{v}_n$ is high (irrelevant of the onset density). 
Our analysis shows that the front velocity achieves its maximum when the shock transition occurs
near the density of the equilibrium PT. At both higher and lower densities the front velocity is smaller than in this case.

In conclusion, we have qualitatively investigated the dynamics of a shock-like discontinuity induced by a PT in a compact star. 
Using the conservation laws at the shock front, we have calculated thermodynamic characteristics in the quark core, and have
estimated the front velocity as a function of the incoming velocity of nuclear matter. This velocity 
is expected to be large when the PT is initiated near the star center.
This should lead to large initial values of the density and pressure jumps. 
We also conclude that the density ratio $\lambda$ depends only weakly on the onset density and pressure jump across the front
as long as the incoming matter velocity is high.
It is shown that the shock wave will first accelerate but then it becomes slower when moving outwards to less dense regions of the star. 
To the best of our knowledge this is a new effect which was not discussed previously.
For a realistic dynamical study of the star transformation one should solve time-dependent hydrodynamical 
equations coupled to the Einstein equation. We hope that our qualitative analysis will be useful to guide such complicated calculations.

The authors are grateful for helpful discussions with Stefan Schramm, Alessandro Brillante and Debades 
Bandyopadhyay. The authors acknowledge financial support from the HIC for FAIR project (Germany). I.M and L.S. acknowledge partial 
support from grant NSH.932.2014.2 (Russia).

\end{document}